\documentclass{elsarticle}

\usepackage{graphicx}
\usepackage[ruled, linesnumbered, vlined, lined, boxed]{algorithm2e}
\usepackage{todonotes}
\usepackage{arydshln}
\usepackage{url}
\usepackage{float}

\usepackage{footnote}
\usepackage{amsmath}

\usepackage[]{hyperref}

\newcommand{\SA}{\ensuremath{\mathsf{SA}}\xspace}

\newcommand{\rank}{\ensuremath{\mathsf{rank}}\xspace}
\newcommand{\select}{\ensuremath{\mathsf{select}}\xspace}
\newcommand{\DA}{\ensuremath{\mathsf{DA}}\xspace}
\newcommand{\BWT}{\ensuremath{\mathsf{BWT}}\xspace}

\newcommand{\EBWT}{\ensuremath{\mathsf{eBWT}}\xspace}

\newcommand{\etal}{{\it et al.}\xspace}

\newcommand{\BM}{\ensuremath{\mathsf{B}}\xspace}
\newcommand{\proc}{\ensuremath{p}\xspace}

\renewcommand{\S}{\mathcal{S}}

\def\gSACAK{{\sf gSACA-K}\xspace}
\newcommand{\BWSD}{\ensuremath{\mathsf{BWSD}}\xspace}

\newcommand{\B}{\ensuremath{\mathsf{\alpha}}\xspace}
\newcommand{\R}{\ensuremath{\mathsf{R}}\xspace}

\newcommand{\DM}{\ensuremath{\mathsf{D_M}}\xspace}
\newcommand{\DE}{\ensuremath{\mathsf{D_E}}\xspace}

\newcommand{\Dm}[2]{\ensuremath{\mathsf{D_M}(#1, #2)}\xspace}
\newcommand{\De}[2]{\ensuremath{\mathsf{D_E}(#1, #2)}\xspace}

\newcommand{\Md}{\ensuremath{\mathsf{M}}\xspace}

\newcommand{\RMQ}{\ensuremath{\mathsf{RMQ}}\xspace}

\newcommand{\rmq}{\ensuremath{\mathsf{rmq}}\xspace}
\newcommand{\Prev}{\ensuremath{\mathsf{prev}}\xspace}
\newcommand{\Next}{\ensuremath{\mathsf{next}}\xspace}

\renewcommand{\tt}{\texttt}

\newcommand{\st}{\texttt{SF}\xspace}
\newcommand{\bit}{\texttt{BIT}\xspace}
\newcommand{\bitsd}{\texttt{BIT\_sd}\xspace}
\newcommand{\wt}{\texttt{WT}\xspace}
\newcommand{\gog}{\texttt{RMQ\_light}\xspace}
\newcommand{\gognz}{\texttt{RMQ}\xspace}

\newtheorem{definition}{Definition}

\begin{document}

\begin{frontmatter}

\title{Algorithms to compute the Burrows-Wheeler Similarity Distribution\tnoteref{mytitlenote}}

\tnotetext[mytitlenote]{
A preliminary version of this work appeared in SPIRE 2018~\cite{Louza2018a}.
}


\author[usp]{Felipe A. Louza\corref{mycorrespondingauthors}}
\cortext[mycorrespondingauthors]{Corresponding author}
\ead{louza@usp.br}
\author[unicamp]{Guilherme P. Telles}
\ead{gpt@ic.unicamp.br}
\author[ebay]{Simon Gog}
\ead{sgog@ebay.com}
\author[usp]{Liang Zhao}
\ead{zhao@usp.br}

\address[usp]{Department of Computing and Mathematics, University of S\~ao Paulo, \\Ribeir\~ao Preto, Brazil}
\address[unicamp]{Instituto de Computa\c{c}\~ao, Universidade Estadual de Campinas, Campinas, Brazil}
\address[ebay]{eBay Inc., San Jose, USA}

\begin{abstract}
The Burrows-Wheeler transform (BWT) is a well studied text transformation widely used in data compression and text indexing. The BWT of two strings can also provide similarity measures between them, based on the observation that the more their symbols are intermixed in the transformation, the more the strings are similar. In this article we present two new algorithms to compute similarity measures based on the BWT for string collections. In particular, we present practical and theoretical improvements to the computation of the Burrows-Wheeler similarity distribution for all pairs of strings in a collection. Our algorithms take advantage of the BWT computed for the concatenation of all strings, and use compressed data structures that allow reducing the running time with a small memory footprint, as shown by a set of experiments with real and artificial datasets.
\end{abstract}


\end{frontmatter}


\section{Introduction}

{
Comparing strings is one of the most fundamental tasks in
Bioinformatics and Information Retrieval~\cite{Ohlebusch2013,Makinen2015,Baeza2011}.
While there exist many measures of similarity between strings,
alignment-based measures are widely used in Bioinformatics because they
are very good in capturing the conservation of blocks of DNA and protein
sequences.  They are, however, computationally intensive to evaluate.
With current databases of biological sequences at the order of hundreds
of gigabytes, alternatives have been proposed both as faster, heuristic
algorithms and as easier to compute similarity measures~\cite{Pizzi16,ThankachanCLKA17,BelazzouguiC17,NojoomiK17,LinAJJ18}.

The Burrows-Wheeler transform (BWT)~\cite{Burrows1994} is a reversible
transformation of a string that tends to group identical symbols into runs
by exploiting context regularities. The intuition of using the BWT as a
means to evaluate distance between strings $S_1$ and $S_2$ is that the more
the symbols in the concatenation of $S_1$ and $S_2$ are intermixed by the
transformation, the greater the number of shared substrings and the more
similar $S_1$ and $S_2$ are.

A class of similarity measures was defined by Mantaci
\etal~\cite{Mantaci2008b} over an extension of the Burrows-Wheeler
transform for string collections, called \EBWT~\cite{Mantaci2005}.
Later, Yang \etal~\cite{Yang2010a,Yang2010b} recrafted the method by
Mantaci \etal and introduced the Burrows-Wheeler similarity distribution
(\BWSD) of two strings $S_1$ and $S_2$ based on the BWT of their
concatenation.  The authors evaluated similarity measures based on the
expectation and Shannon entropy of the \BWSD to efficiently construct
phylogenetic trees for DNA and protein sequences, thus contributing to an
alternative to alignment-based similarity measure among biological
sequences.
}

In this article we present two new algorithms to compute the
Burrows-Wheeler similarity distribution and we show how to
efficiently compute \BWSD-based distances among all pairs of strings in a
collection.  Our algorithms compute the BWT for the concatenation of all
strings only once, instead of the pairwise construction of BWTs proposed by
Yang~\cite{Yang2010a,Yang2010b}, and use compressed data structures that
allow reductions of the running time while still keeping a small memory
usage, as shown by a set of experiments with real and artificial datasets.
We also present both space-efficient alternatives and parallel versions of our
algorithms, that achieved good time/space trade-off and
speedup factors in our experiments, thus
enabling the evaluation of the measure at larger scales.


This article is organized as follows.
Section~\ref{s:background} introduces concepts and notations.
Section~\ref{s:bwsd} presents the \BWSD and their similarity measures.
Sections~\ref{s:alg1} and~\ref{s:alg2} describe our algorithms
with theoretical analysis and implementation alternatives.
Section~\ref{s:experiments} presents experimental results and
Section~\ref{s:conclusion} concludes the article.

\section{Background}\label{s:background}

Let $S[1,n]$ be a string of length $|S|=n$ over an ordered alphabet
$\Sigma$ of size $\sigma$.  The $i$-th symbol of $S$ is denoted by $S[i]$,
with $1\leq i \leq n$.  The substring $S[i] \ldots S[j]$ is denoted by
$S[i,j]$, for $1\leq i \leq j\leq n$.  $S[i,n]$ is the suffix of $S$ that
starts at position $i$.  We assume that $S[n] = \$$ is a terminator symbol
which is not present elsewhere in $S$ and precedes every other symbol in
$\Sigma$.  Juxtaposition is the concatenation operator of strings or
symbols.

\subsection{Suffix array and BWT}

The {\em suffix array} (\SA)~\cite{Manber1993,Gonnet1992} of a string
$S[1,n]$ is an array of integers in the range $[1,n]$ that gives the
lexicographic order of all suffixes of $S$ such that $S[\SA[1], n] <
S[\SA[2], n] < \ldots < S[\SA[n], n]$.  The suffix array may be constructed
in $O(n)$ time using $O(\sigma \lg n)$ bits of workspace~\cite{Nong2013},
which is optimal for strings from constant size alphabets.


The {\em Burrows-Wheeler transform} (BWT)~\cite{Burrows1994} of a string
$S$ is a reversible transformation that tends to group identical symbols into
runs.  It is constructed by sorting the $n$ circular shifts
(conjugates) of $S$, aligning them columnwise and taking the last
column as the \BWT.
Alternatively, the BWT may be obtained concatenating the symbols of $S$
that precede each suffix in the lexicographical order.  Therefore, the BWT
may be defined in terms of the suffix array of $S$, such that

\begin{equation}\label{eq:bwt}
\BWT[i]=
\begin{cases}
S[\SA[i]-1] & \mbox{ if }\SA[i]\neq 1\\
\$ & \mbox{ otherwise.}
\end{cases}
\end{equation}

We define the context $i$ of the BWT as the prefix of the $i$-th sorted
suffix up to and including the terminal symbol $\$$.  The BWT can be
obtained from $S$ and \SA (Equation~\ref{eq:bwt}) or it can be computed directly, without
computing \SA, in $O(n)$ time~\cite{Okanohara2009}
using $O(n \lg \sigma)$ bits of workspace~\cite{MunroNN17}.

The BWT is a well studied text transformation and it is at the heart of many recent advances in
string processing
(see~\cite{Adjeroh2008,Ohlebusch2013,Makinen2015,Navarro2016}).  The
grouping effect of the BWT is used to improve data
compression~\cite{MantaciRRSV17}.  It is also important to the construction
of efficient compressed indices for
strings~\cite{Ferragina2005,Navarro2007}.

Figures~\ref{fig:bwt}(a) and~\ref{fig:bwt}(b) show the BWTs and the contexts
for $S_1=\texttt{banana\$}$ and $S_2=\texttt{anaba\$}$.

\begin{figure}[t]
\begin{center}
\begin{tabular}[t]{ccc}
(a)
&
(b)
&
(c)
\\
\begin{tabular}[t]{|r|c|l|}\hline
$i$& \BWT &  context    \\
\hline
1  &  \tt{a}               & \tt{\$}        \\
2  &  \tt{n}               & \tt{a\$}       \\
3  &  \tt{n}               & \tt{ana\$}     \\
4  &  \tt{b}               & \tt{anana\$}   \\
5  &  \tt{$\$$}                                          & \tt{banana\$}  \\
6  &  \tt{a}               & \tt{na\$}      \\
7  &  \tt{a}               & \tt{nana\$}    \\
\hline
\end{tabular}
&
\begin{tabular}[t]{|r|c|l|}\hline
$i$& \BWT &  context    \\
\hline
1  &  \tt{a}                & \tt{\$}        \\
2  &  \tt{b}                & \tt{a\$}       \\
3  &  \tt{n}                & \tt{aba\$}     \\
4  &  \tt{$\$$}                                                 & \tt{anaba\$}   \\
5  &  \tt{a}                & \tt{ba\$}  \\
6  &  \tt{a}                & \tt{naba\$}      \\
\hline
\end{tabular}
&
\begin{tabular}[t]{|r|c|c|l|}
\hline
$i$ & $\DA$     & $\BWT$ & context \\
\hline
 1  & 1         & \tt{a}           & \tt{$\$_1$}       \\
 2  & 2         & \tt{a}           & \tt{$\$_2$}       \\
 3  & 1         & \tt{n}           & \tt{{a}\$$_1$}       \\
 4  & 2         & \tt{b}           & \tt{{a}\$$_2$}       \\
 5  & 2         & \tt{n}           & \tt{aba$\$_2$}     \\
 6  & 1         & \tt{n}           & \tt{{ana}$\$_1$}     \\
 7  & 2         & \tt{$\$_1$}      & \tt{{ana}ba$\$_2$}   \\
 8  & 1         & \tt{b}           & \tt{{ana}na$\$_1$}  \\
 9  & 2         & \tt{a}           & \tt{{ba}$\$_2$}      \\
 10 & 1         & \tt{$\$_2$}      & \tt{{ba}nana$\$_1$}  \\
 11 & 1         & \tt{a}           & \tt{{na}$\$_1$}    \\
 12 & 2         & \tt{a}           & \tt{{na}ba$\$_2$}      \\
 13 & 1         & \tt{a}           & \tt{{na}na$\$_1$}    \\
\hline
\end{tabular}
\end{tabular}
\end{center}
\caption{BWTs for $S_1=\texttt{banana\$}$, $S_2=\texttt{anaba\$}$ and $S^{cat}=S_1S_2=\tt{banana}\$_1\tt{anaba}\$_2$.}
\label{fig:bwt}
\end{figure}

\subsection{String collections}

Let $\S = \{S_1, S_2, \dots, S_d\}$ be a collection of $d$ strings of
lengths $n_1, n_2, \dots, n_d$ over an alphabet $\Sigma$.  The total
length of $\S$ is $N=\sum_{i=1}^{d}n_i$.  The suffix array for collection
$\S$ can be obtained by computing the \SA of the concatenated string
$S^{cat}[1,N] = S_1 S_2 \ldots S_d$, such that each terminal symbol is
replaced by a symbol $\$_i$, with $\$_i<\$_j$ iff $i<j$.
The BWT for collection $\S$ can be also obtained by the \SA of the
concatenated string as in Equation~\ref{eq:bwt}.

The suffix array of $S^{cat}[1,N]$ is commonly accompanied by the {\em document
array} (\DA), where $\DA[i]$ stores the index of the string which context
$\SA[i]$ came from.
Figure~\ref{fig:bwt}(c) shows the BWT, the document array and the contexts for
$S^{cat}=S_1S_2=\tt{banana}\$_1\tt{anaba}\$_2$.

The suffix array for $\S$ may be constructed in optimal $O(N)$ time
using $O(\sigma \lg N)$ workspace
on $S^{cat}$
without replacing the terminators by distinct symbols
and, as consequence, without increasing the alphabet size,
while still preserving the order
among equal contexts~\cite{Louza2017c}.
The document array for $\S$ can be computed in $O(N)$ time using $O(1)$
workspace along the construction of the suffix array for
$S^{cat}[1,N]$~\cite{Louza2017c}.

\subsection{Rank/select queries and RMQ}

A {\em rank query} on a bitvector $B[1,n]$, denoted by $\rank_1(B,i)$, returns
the number of occurrences of bit $1$ in $B[1,i]$.
A {\em select query} on a bitvector $B[1,n]$, denoted by $\select_1(B,i)$, returns
the position of the $i$-th occurrence of bit $1$ in $B[1,n]$.
$B$ can be preprocessed in $O(n)$ time so that rank/select queries are supported
in $O(1)$ time using $o(n)$ bits of additional space~\cite{Munro1996}.

A {\em wavelet tree}~\cite{Grossi2003} for an array $A[1,n]$ with $\sigma$ distinct
symbols supports rank/select queries in $O(\lg \sigma)$ time.
The wavelet tree uses $n \lg \sigma +o(n \lg \sigma)$ bits of space and
can be built in $O(n \lceil \frac{\lg \sigma}{\sqrt{\lg
n}} \rceil)$ time~\cite{Munro2016}.

A {\em range minimum query} (\rmq) on an array $A[1,n]$ returns the smallest value in
a given interval of $A$, that is, $\rmq(i,j) = \min_{i < k \leq j} \{A[k]\}$ for $1
\le i <j \le n$, whereas a {\em range maximum query} (\RMQ)
returns the largest value in a given interval.
The \rmq and \RMQ operations may be solved in constant
time~\cite{GearyRRR04,OhlebuschG09} with a linear time preprocessing using
$2n+o(n)$ bits of space~\cite{Fischer2006}.

\section{Burrows-Wheeler Similarity Distribution}\label{s:bwsd}

The {\em Burrows-Wheeler similarity distribution} (\BWSD) of a pair of strings
$S_1$ and $S_2$
is constructed as follows.  Given the BWT of $S^{cat}=S_1S_2$, we create a
bitvector $\B_{1,2}$ of size $n_1+n_2$ such that $\B_{1,2}[i]=0$ if $\BWT[i]=\$_2$
or $\BWT[i]$ is a symbol from string $S_1$ and $\BWT[i]\neq\$_1$, and
$\B_{1,2}[i]=1$ otherwise.
In other words, $\B_{x,y}[i]=0$ if $\DA[i]=x$, that is, the $i$-th context came from string $S_x$,
and $\B_{x,y}[i]=1$ if $\DA[i]=y$.

The bitvector $\B_{1,2}$ may be represented as a sequence of runs in
the form $r_{1,2} = 0^{k_1}1^{k_2}0^{k_3}1^{k_4}\ldots0^{k_m}1^{k_{m+1}}$, where
$i^{k_j}$ indicates that $i$ repeats $k_j$ times and such that only $k_1$ and
$k_{m+1}$ may be zero.  Note that $|r_{1,2}|=m+1$ is at most
$2 \cdot (\min(n_1,n_2)+1)$.
Let $t_{k_j}$ be the sum of the number of occurrences of $0^{k_j}$ and
$1^{k_j}$ in $r_{1,2}$.
The largest possible value for $k_j$ is $k_{\max}=\max(n_1,n_2)$.
Let $s = t_1+t_2+\ldots+t_{k_j}+\ldots+t_{k_{\max}}$.

\begin{definition}
$\BWSD(S_1, S_2)$ is the probability mass function $P\{k_j=k\}=t_k/s$ for $k=1,2,\dots, k_{\max}$.
\end{definition}

For example, given strings $S_1= \mathtt{banana\$_1}$ and $S_2= \mathtt{{anaba\$_2}}$
shown in Figure~\ref{fig:bwt}, we have
\[
\begin{array}{c}
\BWT(S_1S_2) = \mathtt{a{a}n{bn}n{\$_1}b{a}{\$_2}a{a}a} \\ 
\B_{1,2}=\{0,1,0,1,1,0,1,0,1,0,0,1,0\} \\ 
r_{1,2} = 0^11^10^11^20^11^10^11^10^21^10^11^0
\end{array}
\]
Therefore, $t_1=9$, $t_2=2$ and $s=11$. The $\BWSD(S_1,S_2)$ is
\[
P\{k_j=1\}=9/11, P\{k_j=2\}=2/11.
\]

Yang \etal~\cite{Yang2010a,Yang2010b} defined the following similarity measures based on the \BWSD
to compare $S_1$ and $S_2$.

\begin{definition}\label{d:expectation}
$\Dm{S_1}{S_2}=E(k_j)-1$, where $E(k_j)$ is the expectation of $\BWSD(S_1,S_2)$.
\end{definition}

\begin{definition}
$\De{S_1}{S_2}=-\sum_{k\geq1, t_k\ne 0}(t_k/s)\lg (t_k/s)$ is the Shannon entropy of $\BWSD(S_1,S_2)$.
\end{definition}

We remark that if $S_1$ is equal to $S_2$, then the \BWSD is
$P\{k_j=1\}=\frac{n_1+n_2}{n_1+n_2}=1$ and $\Dm{S_1}{S_2}=\De{S_1}{S_2}=0$.
Also, note that $\B_{1,2}$ is equal to the complement of $\B_{2,1}$, then both
have the same distribution and $\De{S_1}{S_2}=\De{S_2}{S_1}$ and
$\Dm{S_1}{S_2}=\Dm{S_2}{S_1}$ for any two strings.

\subsection{Straightforward algorithm: $O(dN)$ time}\label{s:yang}

The {\em Burrows-Wheeler similarity distribution} of $S_1$ and $S_2$ 
can be computed straightforward~\cite{Yang2010a,Yang2010b} by first building the BWT of $S^{cat}=S_1S_2$
and the bitvector $\B_{1,2}$, then obtaining $t_1,t_2,\ldots,t_{k_{\max}}$ and
$s$.
The BWT and $\B_{1,2}$ may be constructed in linear time and computing
$t_{k_j}$ also takes linear time.
Therefore computing
$\BWSD(S_1,S_2)$ takes $O(n_1+n_2)$ time. \DM and \DE can be computed in
$O(\max(n_1,n_2))$ time.

Given a collection of $d$ strings of total length
$N=n_1+n_2+\dots+n_d$, computing the matrix $\Md_{d\times d}$ with all pairs of distances (upper
triangular matrix) will take $\sum_{i=1}^{d}\sum_{j>i}^{d}O(n_i+n_j) =
O(dN)$ time.

\section{Algorithm 1: $O(dN)$ time}\label{s:alg1}

Algorithm 1 concatenates all strings into $S^{cat}=S_1S_2\ldots S_d$.  Then it
computes the \BWT and the document array \DA of $S^{cat}$.  In the sequel, the
algorithm builds $d$ bitvectors $\BM_i[1,N]$, where $\BM_i[j]=1$ if
$\DA[j] = i$ or $\BM_i[j]=0$ otherwise, and builds an $O(1)$ rank/select data structure
over each $\BM_i$.
The algorithm then proceeds line by line on the matrix $\Md_{d\times d}$.
To evaluate the distances among $S_i$ and $S_{j>i}$, the algorithm selects the
intervals over $\DA[1,N]$ that contain consecutive occurrences of $i$.
For each interval $[q_s, q_e]$ the algorithm counts the $k_j$ occurrences of $j$, which
corresponds to the existence of the run $0^11^{k_j}0^1$ in the sequence of runs $r_{i,j}$ for $S_i$
and $S_j$.
The runs $0^{\ell_j+1}$ are computed whenever $\ell_j$ consecutive intervals of $i$ do
not contain any occurrence of $j$.
We select the intervals by performing select queries of $\BM_i[1, N]$,
and we count the occurrences of $j$ by performing rank queries over $\BM_j[q_s, q_e]$.

\paragraph{Algorithm 1}
The pseudocode is shown in Algorithm~\ref{a:alg1}.
At each step $i=1, \dots, d$ (Line~3), the algorithm computes the distances in line $i$
of $\Md_{d\times d}$ (Line~25).
Initially, all counters $t^{j}_{k_j}=0$ (Line~4),
$\ell_j=1$ for all $j \in [i+1, d]$ (Line~5),
and $q_s=1$ (Line~6).

Given a collection of strings $\S = \{S_1, S_2 \dots, S_d\}$ as input,
Algorithm 1 outputs a strictly upper triangular matrix $\Md_{d\times d}$, where each entry
$\Md[i][j]$ is either \Dm{S_i}{S_j} or \De{S_i}{S_j}.
For $p=1,\dots,n_i$ (Line~7), the algorithm sets $q_{e}$ such that $\DA[q_{e}]$
corresponds to the $p$-th value equal to $i$ in $\DA[1,N]$ (Line~8).
At the end of the iteration, $q_{s}$ receives $q_e$ (Line 19).

Then, given the current interval $\DA[q_{s},q_e]$,
for each  $j \in [i+1, d]$ (Line~9), it counts the number of $j$'s in the interval
by computing $\rank_1(\BM_j,q_e)-\rank_1(\BM_j,q_s)$ and stores it
in $k_j$ (Line~10).
If $k_j>0$ it means that the run $0^{\ell_j}1^{k_j}$ occurs in
$r_{i,j}$, thus $t^j_{k_j}$ and $t^j_{\ell_j}$ are increased by 1 and $\ell_j$
gets 1 for the next iteration (Lines~12-14).
Otherwise, if $k_j=0$, it means that the block $0^{\ell_j}1^00^1$ in $r_{i,j}$
may be collapsed into $0^{{\ell_j+1}}$ in a next iteration.
To this end, $\ell_j$ must be increased by one (Line 16).
Then, when $k_j>0$ or when the algorithm reaches Line 23,
counter $t^j_{\ell_{j}}$ is increased by one.

The select queries over $\BM_i[1,N]$ (Line 8) will enable selecting up to the
last symbol $\DA[q_e]=i$, but there can be more symbols in $\DA[q_e,N]$
equal to $j$, for all $j>i$.
Then, Lines 22-24 deal with the last blocks of 0s and 1s accordingly and
invoke the computation of the distance measure from the counters $t^j$ (Line 25).

\begin{algorithm}[]
\renewcommand{\to}{\mbox{\bf\xspace to \xspace}}
\SetNlSty{textbf}{}{}
\SetAlgoLined
\SetCommentSty{mycommfont}
\KwData{$\S = \{S_1, S_2, \dots, S_d\}$, $|S_i|=n_i$}
\KwResult{result matrix $\Md_{d\times d}$}

Build $\BWT$ and $\DA$ for $S^{cat}=S_1S_2\dots S_d$

Compute $\BM_i$, for $i=1,2,\dots,d$;

\For{$i \leftarrow 1 \to d$}{

    $t^{j}_{k_j} \leftarrow 0$ for all $k_{j}$;

    $\ell_j \leftarrow 1$ for all $j>i$;

    $q_s \leftarrow 1$;

    \For{$p \leftarrow 1 \to n_i$}{

                        $q_e \leftarrow \select_1(\BM_i, p)$;

                        \For{$j \leftarrow i+1 \to d$}{

                                $k_j \leftarrow \rank_1(\BM_j,q_e)-\rank_1(\BM_j,q_s)$;

                                \eIf{$k_j>0$}{
                                        $t^{j}_{k_j}$\tt{++} \tcp*[r]{$1^{k_{j}}$}
                $t^{j}_{\ell_{j}}\tt{++}$ \tcp*[r]{$0^{\ell_{j}}$}
                $\ell_{j} \leftarrow 1;$
        }
        {
          $\ell_{j}\tiny{++};$
        }

      }
      $q_s \leftarrow q_e$;
    }

    \For{$j \leftarrow i+1 \to d$}{

     $k_j \leftarrow \rank_1(\BM_j,N)-\rank_1(\BM_j,q_s)$;

             $t^{j}_{k_j}\tt{++}$ \tcp*[r]{$1^{k_{j}}$}

      $t^{j}_{\ell_{j}}\tt{++}$ \tcp*[r]{$0^{\ell_{j}}$}

       $\Md[i][j] \leftarrow compute\_distance(\tt{$t^{j}$}, n_i, n_j)$; 
        }
}
\caption{Compute Distances}
\label{a:alg1}
\end{algorithm}

We remark that the second rank operation $\rank_1(\BM_j,q_s)$ of Line 10 at
iteration $p$ can be avoided by storing the result of the first rank operation,
$\rank_1(\BM_j,q_e)$, of iteration $p-1$, where $q_e$ was equal to
$q_s$.
The same idea can be applied for Line 22.
Another practical improvement can be achieved by storing, in an auxiliary
array of size $N$, for each position $\DA[i]=j$ the position of next value equal to
$j$ in $\DA[i+1,N]$, such that, in the for loop of Line 9, whenever the next position equal to
$j$ in \DA is greater than $q_e$, we can avoid two rank operations and go directly to
Line 16 (in this case $k_j=0$).

\subsection{Theoretical costs}

\BWT and \DA can be computed in $O(N)$ time using $O(\sigma \lg N)$ bits of
workspace~\cite{Louza2017c}.
The construction of all bitvectors $\BM_i[1,N]$ with rank/select support takes
$O(dN)$ time.  
For each string $S_i$ the algorithm performs $n_i$ select operations (Line 8),
each one in $O(1)$ time, and performs $(n_i+1) d$ rank operations (Lines 10 and
22), each one in $O(1)$ time.
The cost to compute each distance (Line 25) is $O(n_i+n_j)$.
Therefore, the total running time is $O(dN)$ time.

The workspace used by the algorithm is $N\lg\sigma$ bits for $S^{cat}$,
$N\lg\sigma$ bits for the \BWT,
$dN+o(dN)$ bits for the bitvectors, and
$2 \cdot (d \lg( \max(k_j)))$ bits for the lists $k_j$, $\ell_j$,
and $d \cdot (\max(k_j) \lg( \max(k_j)))$ bits to store all counters $t^{j}_{k_j}$, where
$\max(k_j)$ is bounded by longest string length in the collection.
We remark that after computing the bitvectors, the space of \DA can be released.

\subsection{Implementation alternatives}\label{s:alternatives1}

Note that each bitvector $\BM_i[1,N]$ will be very sparse, containing exactly
$n_i$ bits equal to $1$.
We discuss two space-efficient alternatives to reduce the workspace of Algorithm 1.

\paragraph{Sparse bitvectors}
We can use Elias-Fano compressed bitvectors with rank/select support~\cite{Okanohara2007}, such that
each $\BM_i$ will take $n_i \lg \frac{N}{n_i} + 1.92n_i + o(n_i)$ bits of space.
The total space will be reduced to
\begin{align*}
\sum_{i=1}^d( n_i \lg \frac{N}{n_i} + 1.92n_i + o(n_i)) =
\sum_{i=1}^d~(n_i\lg\frac{N}{n_i})~+ 1.92N + o(N) =\\
N \sum_{i=1}^d (\frac{n_i}{N} \lg \frac{N}{n_i}) + 1.92N +o(N) =
N H_0(\DA) + 1.92N + o(N) \mbox{ bits,}
\end{align*}
where $H_0(\DA)$ is the entropy compressed size of \DA.
The running time will increase to $O(d N \lg{\frac{N}{avg(n_i)}})$,
because each rank operation will take $O(\lg{\frac{N}{n_i}})$ time, where $avg(n_i)$ is the average length of the strings.

\paragraph{Wavelet trees}
Another alternative is to replace all bitvectors $\BM_1, \BM_2, \dots, \BM_d$ by a single
wavelet tree built over $\DA[1,N]$.
The alphabet size of such wavelet tree will be equal to $d$.
Therefore, the space used by the $d$ bitvectors will be reduced to $N\lg d + o(N\lg d) \mbox{ bits}$
for the wavelet tree with rank/select support.
On the other hand, the running time will increase to $O(dN\lg d)$, because
each rank and select
operations will take $O(\lg d)$ time.

\subsection{Parallel version}\label{s:par1}

The for loop of Line 3 can be parallelized to compute at the same time all lines of
matrix $\Md_{d\times d}$ using multiple threads.
To this end, each thread may have a local
copy of variables $q_s$, $q_e$, lists $k_{j}$ and $\ell_j$, and
counters $t^{j}_{k_j}$,
while the bitvectors $\BM_1, \BM_2, \dots, \BM_d$ with
rank/select support (or the wavelet tree), and the output matrix $\Md_{d\times d}$ can be shared.
The total running time will be reduced to $O(dN/\proc)$, where \proc is the number
of threads.
On the other hand,
the workspace will increase to $\proc \cdot (2 \cdot d \lg (\max(k_j)))$ bits for the local lists
and $\proc \cdot (d \cdot \max(k_j) \lg N)$ bits for the counters.

\section{Algorithm 2: $O(n+z)$ time}\label{s:alg2}

Given a collection of unsimilar strings $\S = \{S_1, S_2 \dots, S_d\}$ as input,
the number of runs in all $r_{i,j}$, say $z$, is much smaller
than the maximal possible $O(dN)$.
In the extreme case each $\BM_i$ consists of only
three runs {$0^{N_1}1^{n_i}0^{N_2}$}, with $N_1+n_i+N_2=N$, and the sum of all runs is therefore as
small as $z=d^2-d$.
However, Algorithm 1 would still require $O(dN)$ steps to
count all runs in this case. 
We will show how to improve the running time to $O(N+z)$ using the document-listing
solution by Muthukrishnan, 2002~\cite{Muthukrishnan2002} that allow us to find
all $r$ distinct documents in a given interval of \DA in $O(r)$ time.

Algorithm 2 concatenates all strings into $S^{cat}=S_1S_2\ldots S_d$, and
computes the \BWT and the document array \DA of $S^{cat}$.
Then, it computes
the auxiliary arrays $\Prev[1,N]$ and $\Next[1,N]$, such
that $\Prev[i] = \max\{ j | j<i \mbox{ and } \DA[j]=\DA[i] \}$ or $-1$ if no
such $j$ exists, and $\Next[i] = \min\{j | j>i \mbox{ and } \DA[j]=\DA[i]\}$ or
$N$ if no such $j$ exists.
Also, it computes a range minimum query structure on \Prev\ ($\rmq_{\Prev}$) and a range
maximum query structure on \Next\ ($\RMQ_{\Next}$) in order to
extract the leftmost and the rightmost occurrence of all $r$ distinct documents
in any arbitrary range in $\DA[1,N]$ in $O(r)$ time.
Then, computing an array $\R[1,N]$, where
$\R[i]=rank_{\DA[i]}(\DA, i)$, allows to get the frequency of each distinct document
in $O(1)$ time.

The pseudocode is shown in Algorithm~\ref{a:alg2}.
At each step $q_s=1,\dots,N$ (Line 6) the algorithm process
the intervals $[q_s,q_e]$ of consecutive positions of
symbol $\DA[q_s]=i$ (Lines 6-17).
Initially, $i$ receives $\DA[q_s]$ (Line 7) and $q_e$ receives $\Next[q_s]$ (Line 8),
which points to the next position in $\DA[q_s+1,N]$ equal to $i$.

\begin{algorithm}
\renewcommand{\to}{\mbox{\bf\xspace to \xspace}}
\SetNlSty{textbf}{}{}
\SetAlgoLined
\SetCommentSty{mycommfont}
\KwData{$\S = \{S_1, S_2, \dots, S_d\}$, $|S_i|=n_i$}
\KwResult{result matrix $\Md_{d\times d}$}

Build $\BWT$ and $\DA$ for $S^{cat}=S_1S_2\dots S_d$

Compute $R$, $\Prev$ and $\Next$;

Build $\rmq_{\Prev}$, $\RMQ_{\Next}$;

$t^{i,j}_{k_j} \leftarrow 0$ for all $i, j$ and $k_{j}$;

$Stack \leftarrow \emptyset$;

\For{$q_s \leftarrow 1 \to N$}{

    $i \leftarrow \DA[q_s]$;

    $q_e \leftarrow \Next[q_s]$;

    $Stack \leftarrow $ document-listing($q_s, q_e$) \tcp*[r]{Stack pair: $\big<j, a_j, b_j\big>$}

      \While{Stack is not empty}{

      $\big<j, a_j, b_j\big> \leftarrow Stack.pop()$

      $k_j \leftarrow \R[b_j]-\R[a_j]+1$

      $i' \leftarrow \min(i,j)$;

      $j' \leftarrow \max(i,j)$;

      $t^{i',j'}_{k_j}$\tt{++}; 
			}

}

\For{$i \leftarrow 1 \to d$}{
        \For{$j \leftarrow i+1 \to d$}{
                $\Md[i][j] \leftarrow compute\_distance(\tt{$t^{i,j}$}, n_i, n_j)$;
        }
}

\caption{Compute Distances}
\label{a:alg2}
\end{algorithm}

The algorithm solves the {\em document listing} problem using $\rmq_{\Prev}$
and $\RMQ_{\Next}$ to determine all $r$ distinct documents
that occurs in the interval $\DA[q_s, q_e]$ in $O(r)$ time.
For each distinct document $S_j$, it adds to the Stack the tuple $\big< j, a_j,
b_j\big>$ corresponding to $S_j$ and their leftmost and rightmost
positions in the interval $\DA[q_s, q_e]$ (Line 9).
Then, for each tuple in the Stack, it pops $\big< j, a_j, b_j\big>$ and
computes the frequency $k_j$ of values equal to $j$ in $\DA[q_s,q_e]$ using the values in positions $a_j$ and $b_j$ of array $R$ (Line 12).

In order to compute only the upper triangular matrix of $\Md_{d\times d}$, it computes
the minimum $i'$ and maximum $j'$ values between $i$ and $j$ (Lines 13 and 14) and
the counter $t^{i',j'}_{k_j}$ is increased by one (Line 15).
At the end, the algorithm invokes the computation of the distance measure from the counters $t^{i,j}$ for each pair $S_i,S_{j>i}$ (Lines 18-22).

We remark that during step $q_s$, with $\DA[q_s]=i$, it is not necessary
maintaining the counters of runs $0^{\ell_j}$ that correspond to string
$S_i$ as this is calculated symmetrically in a next step when other strings
$\DA[q_s]\not=i$ are traversed.

\subsection{Theoretical costs}
The precomputation of \BWT, \DA, $R$, \Prev, \Next, $\rmq_{\Prev}$, $\RMQ_{\Next}$ and
requires $O(N)$ time and space.
Generating all intervals requires $\sum_{i=1}^d(n_i+1)=O(N)$ time and each run $1^{k_j}$
of every $S_j$ is handled in constant time with $\R[b_j]-\R[a_j]$.
Overall the time complexity is bounded from above by $O(N+z)$,
where $z$ is the sum of all runs of all pairs $r_{i,j}$ ($1\leq i<j\leq d$).
We remark that the cost to compute all pairs of distances given the counter
$t^{i,j}$ is still $O(dN)$.

The workspace used by the algorithm is $2(N\lg\sigma)$ bits for the
strings $S^{cat}$ and \BWT,
$N\lg d$ bits for \DA,
$3 (N\lg N)$ bits for $R$, \Prev and \Next,
$2 (2N+o(N))$~bits for $\rmq_{\Prev}$, $\RMQ_{\Next}$,
$O(d \lg N)$ bits for the stack,
and
$O(d^2 \max(k_j)) \lg N$ bits to store a quadratic matrix with all
counters $t^{i,j}_{k_j}$ for all pair of strings, where $\max(k_j)$ is bounded by
longest string length in the collection.

\subsection{Implementation alternatives}\label{s:alternatives2}

Algorithm 2 uses a quadratic matrix to store the counters $t^{i,j}_{k_j}$ in
memory, which is a clear spot for improvement in this strategy.

\paragraph{Lightweight version}
We can rewrite Algorithm 2 to first compute distances between string $S_i$ and
 $S_{j>i}$ regarding only positions where $\DA[q_s]$ is equal to $i$, for
$i=1,2,\dots,d$.
Therefore, no quadratic matrix structure is
needed because the counters $t^{i,j}_{k_j}$ will always refer to the same $S_i$
in iteration $i$ and we can replace them by $t^{j}_{k_j}$ as in
Algorithm 1.
However, the theoretical running time will increase to $O(dN)$ because we have
to scan $\DA[1,N]$ $d$ times.

\subsection{Parallel version}\label{s:par2}

The for loop of Line 6 can be parallelized to compute all lines of
matrix $\Md_{d\times d}$ using multiple threads.
Again, each thread may have a local copy of variables $q_s$, $q_e$, $i$, list $k_j$ and a local Stack.
The matrix of the counters $t^{i,j}$ can be shared with locks on writing operations (Line 15).
The arrays $R$, \Prev, \Next and \DA, the \RMQ and \rmq data structures, and the
output matrix $\Md_{d\times d}$ can be shared.
The total running time will be reduced to $O((N+z)/\proc)$, where \proc is
the number of threads.
The workspace will increase to $\proc \cdot (d \lg (\max(k_j)))$ bits for the local lists
and $\proc \cdot O(d \lg N)$ bits for the local stacks.

\section{Experiments}\label{s:experiments}

We have analyzed the performance of the algorithms for
computing the upper triangular entries of matrix $\Md_{d\times d}$.
We computed the expectation based distance \DM (Definition~\ref{d:expectation}).
We compared the straightforward approach (\st)
by Yang \etal~\cite{Yang2010b} with
three versions of Algorithm 1, using plain bitvectors
(\bit), using Elias-Fano compressed bitvectors (\bitsd) and using a wavelet
tree (\wt), and with Algorithm 2 (\gognz) and its {\em lightweight} version (\gog).
We also evaluated the performance of all algorithms running in parallel, in a
shared-memory multithreading environment.

The algorithms were implemented in C++ using the SDSL library~\cite{Gog2014a}
version 2.0\footnote{\url{https://github.com/simongog/sdsl-lite}}.
The parallel versions were implemented using C++ OpenMP.
The BWTs and document arrays were computed with algorithm
\gSACAK\footnote{\url{https://github.com/felipelouza/gsa-is/}}~\cite{Louza2017c}.
The source code of all algorithms is freely available at
\url{https://github.com/felipelouza/bwsd}.

The experiments were conducted on a machine with GNU/Linux 64 bits
operating system (Debian 8, kernel 3.16.0-4) with an Intel Xeon processor
E5-2630 v3 20M Cache 2.40-GHz, 384 GB RAM and 13 TB SATA storage.
The sources were compiled by g++ v 4.9.2, with flags
\texttt{std=c++14}, \texttt{-O3}, \texttt{-m64} and
\texttt{-fomit-frame-pointer}.

We used four different real data collections with up to $d=$15,000 strings,
described in Table~\ref{t:data}.

\begin{savenotes}
\begin{table}[ht]
\centering
\caption{
Datasets used in our experiments. 
Column $2$ reports the alphabet size.
Column $3$ reports the collection size.
Column $4$ shows the number of strings.
Column $5$ and $6$ report the maximum and average length of each string.
}
\setlength{\tabcolsep}{4pt}
\renewcommand{\arraystretch}{1.2}
\label{t:data}
\begin{tabular}{lrrrrr}
\hline
dataset   & $\sigma$ & total length & n. of strings & max length & avg length  \\
\hline
{\sc reads}     & 4        & 1,422,718   & 15,000               & 101                     & 94.85    \\
{\sc uniprot}   & 25       & 3,454,210   & 15,000               & 2,147           & 230.28   \\
{\sc ests}      & 4        & 11,313,165  & 15,000       & 1,560           & 754.21   \\
{\sc wikipedia} & 208      & 25,430,657  & 15,000       & 150,768               & 1,695.38 \\
\hline
\end{tabular}

\begin{description}
\item[{\sc reads}:] is a collection of reads from Human Chromosome 14
  (library 1)\footnote{\url{http://gage.cbcb.umd.edu/data/index.html}}.

\item[{\sc uniprot}:] is a collection of protein sequences from
  Uniprot/TrEMBL protein database release
  2015\_09\footnote{\url{http://www.ebi.ac.uk/uniprot/download-center/}}.

\item[{\sc ests}:] is a collection of DNA sequences of ESTs from {\it
  C. elegans}\footnote{\url{http://www.uni-ulm.de/in/theo/research/seqana.html}}.

\item[{\sc wikipedia}:] is a collection of pages from a snapshot of the
  English-language edition of
  Wikipedia\footnote{\url{http://algo2.iti.kit.edu/gog/projects/ALENEX15/collections/ENWIKIBIG/}}.
\end{description}
\end{table}
\end{savenotes}

\subsection{Running time}

\begin{figure}
 \centering
 \includegraphics[width=1.0\textwidth]{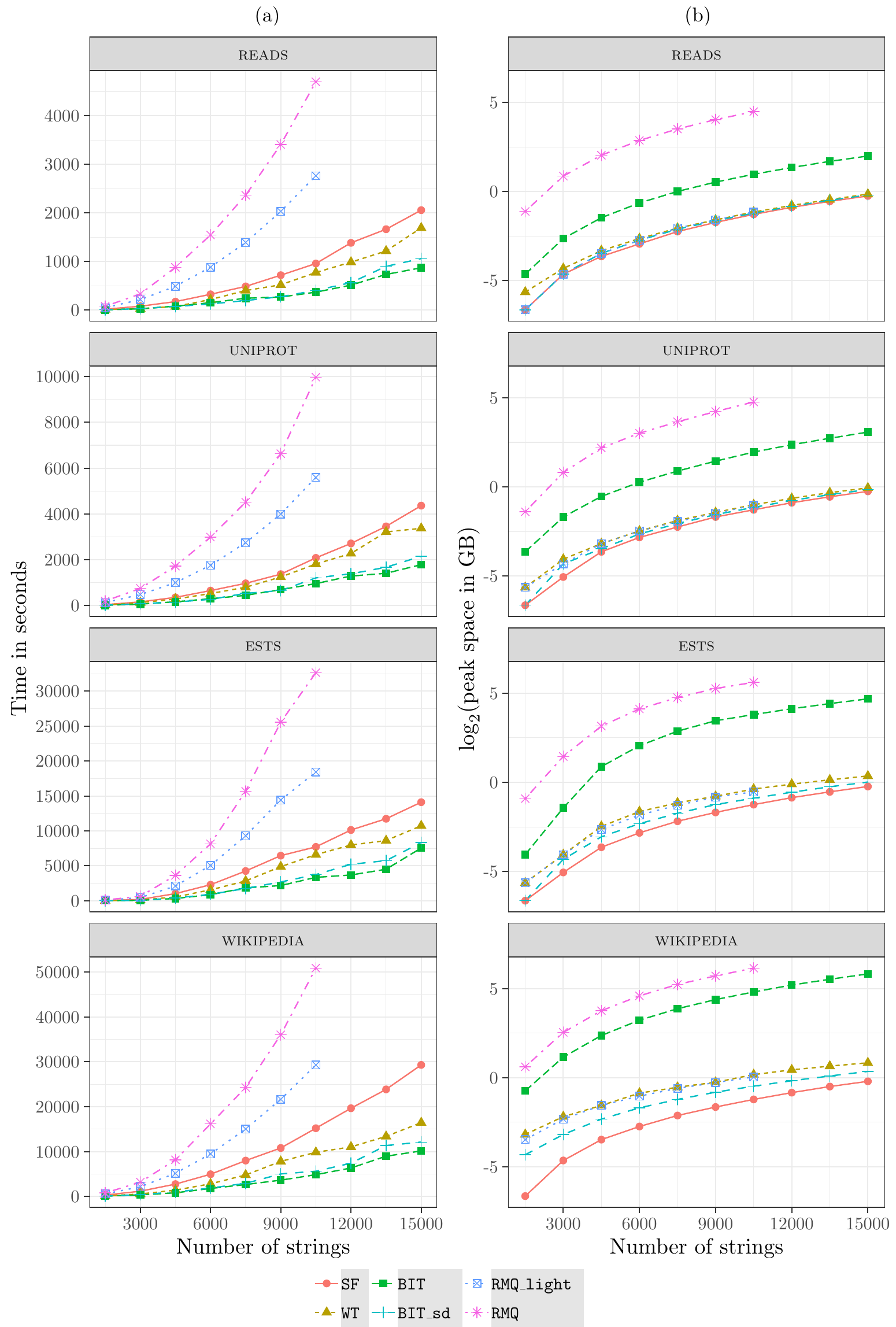}
\caption{Running time in seconds and peak memory in GB (in logarithmic
  scale) for the alternatives of Algorithms 1 and 2 and for the straightforward
  algorithm on all datasets.}
\label{f:all}
\end{figure}

Figure~\ref{f:all}(a) shows the running time in seconds of the algorithms,
measured using the \texttt{clock()} function of ANSI-C.
The running time includes the time spent in building all auxiliary data
structures, which is less than $1\%$ of the total time.
We stopped the execution of \gognz and \gog at $d=10,500$ strings, since it
was clear that its running time was going to exceed the others by far.

\bit and \bitsd were the fastest in all experiments.  Comparing
with the straightforward algorithm, \bit was $2.4$ times faster than \st
while \bitsd was $2.0$ times faster than \st, on the average.
For \textsc{wikipedia}, \bit was $2.9$ times faster than \st, whereas \bitsd
was approximately $2.4$ times faster.
\wt was $1.4$ times faster than \st,
on the average.
On the other hand, \st was $4.32$ times faster than \gognz,
and \st was $2.47$ times faster than \gog, on the average.
In Section~\ref{s:worst} we will discuss an unlikely case where
the performance of \gognz is better.

This results support Algorithm 1 as a practical improvement for computing
matrix $\Md_{d\times d}$, even with the additional time taken by the
rank/select operations when plain bitvectors (\bit) are replaced by
compressed bitvectors (\bitsd) or wavelet trees (\wt).
Algorithm~1
performed better than the \st and than Algorithm~2 on
all inputs.

\subsection{Peak memory}\label{s:peak}

Figure~\ref{f:all}(b) shows the $\log_2$ peak memory usage in GB of each
algorithm measured by the malloc\_count
library\footnote{\url{http://panthema.net/2013/malloc\_count}}.
We remark that the
input collection uses $N$ bytes, whereas the output matrix takes
$(d^2-d)/2$ entries (upper triangular matrix), each one of $8$ bytes
(double variable).  The total size of the output matrix was
approximately 868 MB for collections with 15,000 strings.

The space used by \st was the smallest.  As expected it was very close to
what is needed for the input and output, as only $O(2 \cdot \max(n_i))$ bytes are
added for each pair of the \BWT and auxiliary variables.
The implementations \bitsd, \wt and \gog were also lightweight.
For \textsc{wikipedia} with $d=$ 10,500,
\st used approximately $0.43$ GB,
\bitsd used approximately $0.72$ GB,
\wt used approximately $1.13$ GB and
\gog used approximately $1.03$ GB.
The space used by \bit and by \gognz were, however, much larger.
For \textsc{wikipedia}, \bit and \gognz used approximately $64$
and $165$ times more space than \st, respectively.
We remark that the data structures used by all
versions of Algorithm 1 were the same, except for bitvectors and wavelet
tree.

This result shows that the space used by the plain bitvectors (\bit) may be
a bottleneck for Algorithm 1, and the space used by \gognz becomes infeasible.
We may conclude that the compressed data structures used by
\bitsd and \wt provide good space-efficient alternatives comparable to \st
and \gognz.

The experiments support the the conclusion that \bitsd is a good time/space
trade-off of Algorithm~1.

\subsection{Parallel versions}\label{s:parallel}

The
algorithms were parallelized such that each thread solves each line of
matrix $\Md_{d\times d}$ independently.  We used different number of
threads ($1, 2, 4, 8, 16$ and $32$).  We used the first $d=10,500$ strings
of the four data collections described in Table~\ref{t:data}.  The elapsed
time was taken by the directive \texttt{omp\_get\_wtime()}.

\begin{figure}
 \centering
 \includegraphics[width=1.0\textwidth]{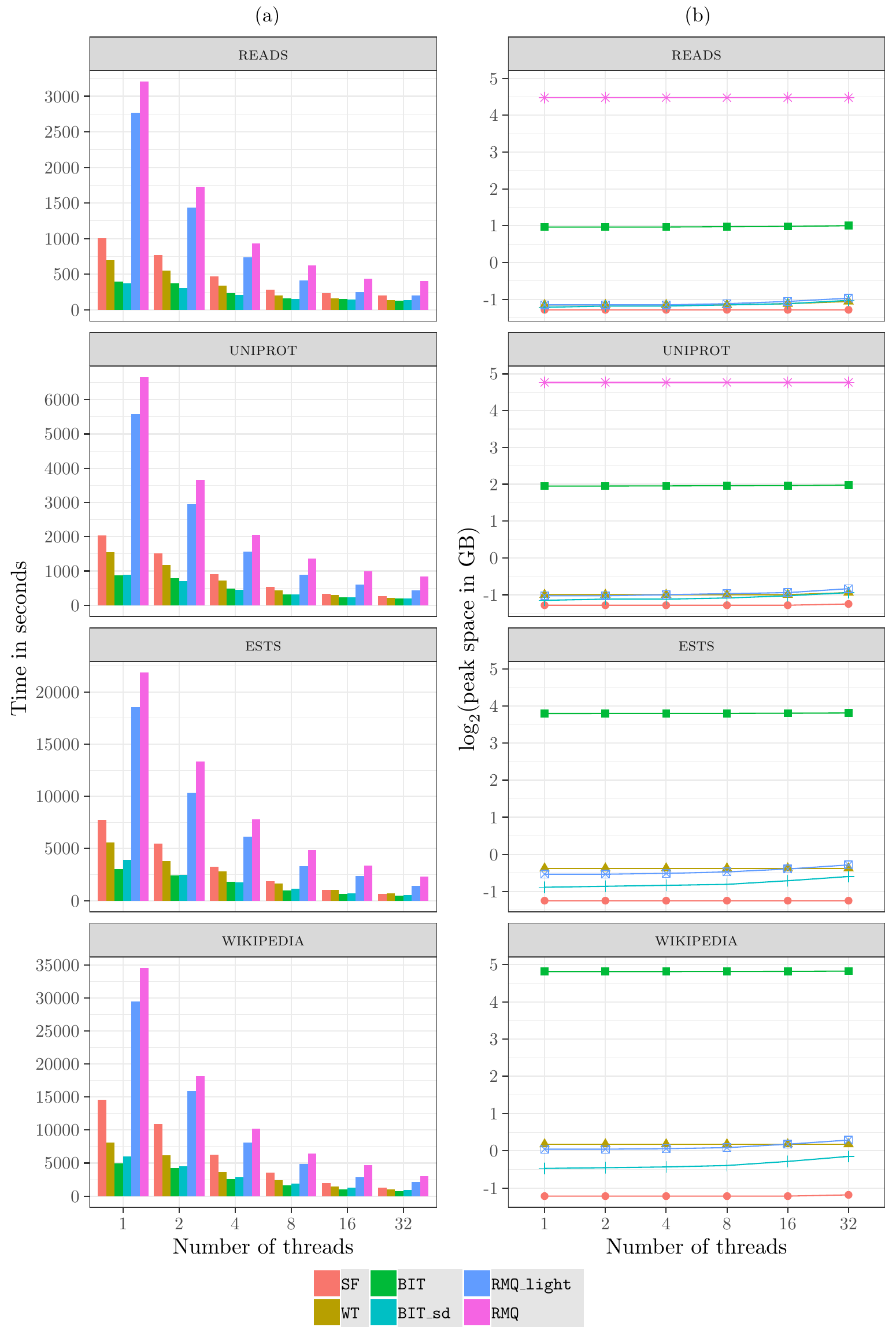}
\caption{Running time in seconds and peak memory in GB (in logarithmic
  scale) for the parallelized alternatives of Algorithms 1 and 2 and for the straightforward
  algorithm on all datasets with $d=10,500$.}
\label{f:parallel}
\end{figure}

Figure~\ref{f:parallel}(a) shows the running time in seconds of each
parallel algorithm as the number of threads increase.
\bit and \bitsd were
still the fastest algorithms with every number of threads.
However, \gog presented a much better speedup with the
increasing number of threads, as shown in
Table~\ref{t:speedup}.

Figure~\ref{f:parallel}(b) shows the $\log_2$ peak memory usage in GB of
each algorithm.  The memory usage increased slightly for all algorithms, due to
local copies of variables and lists used to compute
counters $t^{j}_{k_j}$.

\begin{table}[t]
\setlength{\tabcolsep}{5pt}
\renewcommand{\arraystretch}{1.1}
\centering
\caption{
The table shows the running time in seconds and the speedup of the parallel algorithms over its serial versions, when the numbers of threads is 1.
}
\label{t:speedup}
\begin{tabular}{r|rr|rr|rr}
           & \multicolumn{2}{c|}{\st} & \multicolumn{2}{c|}{\bitsd}                             & \multicolumn{2}{c}{\gog}                               \\
n. threads & \multicolumn{1}{c}{time} & \multicolumn{1}{c|}{speedup} & \multicolumn{1}{c}{time} & \multicolumn{1}{c|}{speedup} & \multicolumn{1}{c}{time} & \multicolumn{1}{c}{speedup} \\ \hline
1  & 1,002.93 &      & \textbf{375.74} &      & 2,769.93 &                \\
2  & 772.56   & 1.30 & \textbf{307.99} & 1.22 & 1,438.67 & \textbf{1.93}  \\
4  & 468.99   & 2.14 & \textbf{210.50} & 1.78 & 735.48   & \textbf{3.77}  \\
8  & 283.75   & 3.53 & \textbf{155.86} & 2.41 & 414.03   & \textbf{6.69}  \\
16 & 233.29   & 4.30 & \textbf{148.57} & 2.53 & 252.44   & \textbf{10.97} \\
32 & 198.29   & 5.06 & \textbf{134.49} & 2.79 & 205.28   & \textbf{13.49}
\end{tabular}
\end{table}

\subsection{Dissimilar strings}\label{s:worst}

We compared all algorithms on an artificial input where
all strings are completely ``different'', for instance when
they come from interleaved and disjoint alphabets.
In a situation like this, the document array is composed by $d$ runs.
We used the dataset \texttt{READS} to compute \BWT and \DA,
then, we artificially replaced the entries of $\DA[1,N]$ such that
$\DA~=~\{1^{N/d}2^{N/d}\dots d^{N/d}\}$.

\begin{figure}
 \centering
 \includegraphics[width=1.0\textwidth]{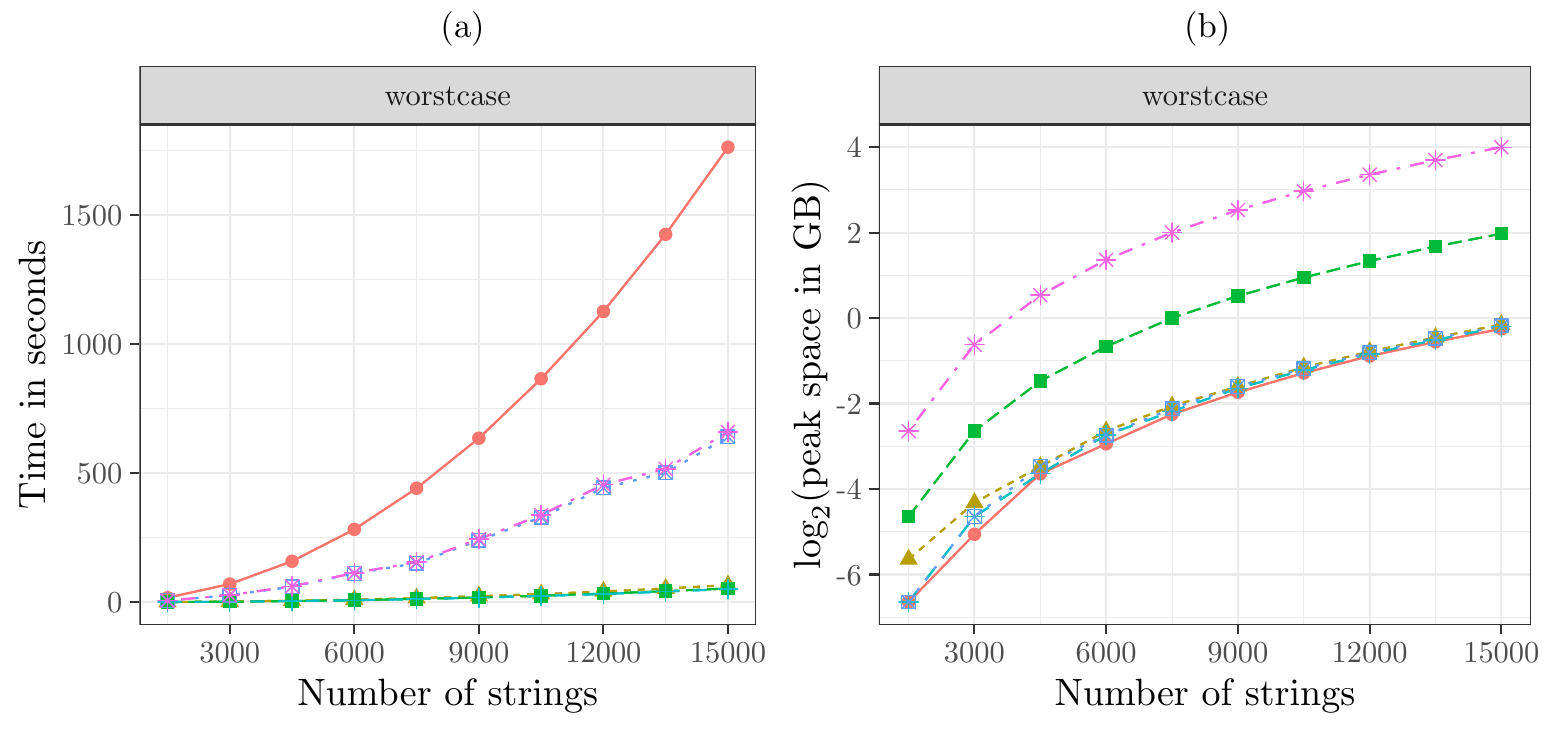}
 \includegraphics[width=1.0\textwidth]{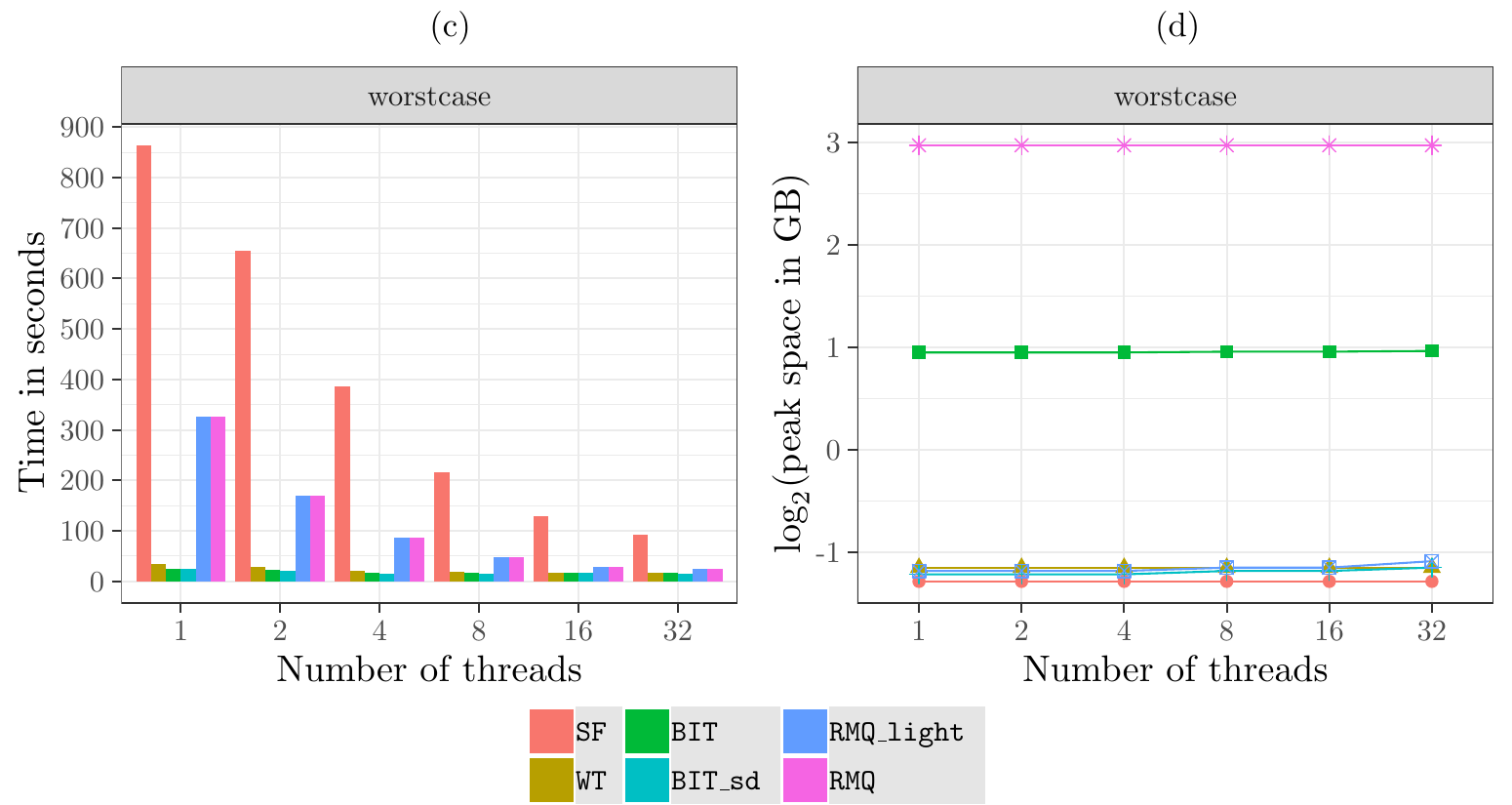}
\caption{Running time in seconds and peak memory in GB (in logarithmic
  scale) for all algorithms on an artificial datasets.}
\label{f:worst}
\end{figure}

Figure~\ref{f:worst}(a) shows the running time in seconds and
Figure~\ref{f:worst}(b) shows the $\log_2$ peak memory usage in GB of each
algorithm.
\bit and \bitsd were again the fastest and \wt was also fast.
Avoiding rank operations in Algorithm 1 greatly influences this result
(see Section~\ref{s:alg1}).
Notice that \gognz and \gog were very close, being $2.75$ times faster than
\st in this experiment, reversing the behavior shown for real datasets. The peak memory
was close to the results obtained in Section~\ref{s:peak}

Figure~\ref{f:worst}(c) shows the running time in seconds
and Figure~\ref{f:worst}(d) shows the $\log_2$ peak memory usage in GB
of each algorithm
running in parallel with $d=10,500$ strings.
\gognz and \gog achieved an
impressive speedup, being faster than \st and getting closer to the
other algorithms.
The results of peak memory were similar to the results obtained previously.

This result shows that the unlikely situation where all strings are completely
``different'', the performance of Algorithm 2 may pay off.

\section{Conclusions}\label{s:conclusion}

In this article we have presented two new algorithms to calculate the
Burrows-Wheeler similarity distribution for all pairs of strings in a
collection.
Our algorithms take advantage of the BWT computed for the concatenation of all
strings.
Algorithm~1 is based on using rank queries on bitvectors or on a
wavelet tree, and Algorithm~2
is based on solving the document-listing problem.  We have also explored
optimized and parallel implementation variants of the algorithms.

The algorithms were analyzed by experiments on a set of real and artificial
collections of strings, having the straightforward algorithm that builds a
BWT for each pair of strings as a baseline.  The experiments revealed a
wide picture of our algorithms' behavior.  Three different versions of
Algorithm~1 outperformed the straightforward algorithm by a factor of up to
$2.9$.  Two versions of our algorithms exhibited a small memory footprint.
Moreover, we obtained good scalability with our parallel variants.

Our algorithms contribute for solving string comparison problems in
practice and are specially interesting for the case of biological sequences
and other large datasets.  While building large phylogenies or comparing a
sequence against a large dataset, like current databases of biological
sequences, the parallel variants may be quite useful.  Other types of
applications may benefit as well, for instance, when investigating
relations among textual documents through visual
phylogenies~\cite{2011-paiva}.

The algorithms we presented here may be extended to evaluate different
similarity measures as well, broadening their application and enabling the
definition of a class of measures of similarity among strings that is also
feasible in practice for large datasets.

\section*{Acknowledgments}
The authors thank Prof. Nalvo Almeida for granting access to the machine used
for the experiments.

\paragraph{Funding}
F.A.L. was supported by the grant $\#$2017/09105-0 from the S\~ao Paulo
Research Foundation (FAPESP).
G.P.T. acknowledges the support of CNPq grants 425340/2016-3 and 310685/2015-0.


\end{document}